\documentclass[twocolumn,floatfix,prd,authoryear,nofootinbib,aps,10pt,tightlines,eqsecnum]{revtex4-2}
\usepackage{bm}
\usepackage{newtxmath,newtxtext}
\usepackage{graphicx}
\usepackage{textcomp}
\usepackage{float}
\usepackage{booktabs}
\usepackage{dcolumn,enumerate}
\usepackage{ragged2e}
\usepackage{hyperref}
\usepackage{caption}

\hypersetup{colorlinks=true, linkcolor=blue, urlcolor=blue, citecolor=blue}

\usepackage{xcolor}
\usepackage{epsfig}
\usepackage{caption}
\usepackage{appendix}
\usepackage{subcaption}
\usepackage{graphicx} 
\usepackage{orcidlink}
\usepackage[dvipsnames]{xcolor}

\begin{document}
\title{Periodic Orbits and Their Gravitational-Wave Signatures around a Rotating Black Hole Surrounded by Perfect Fluid Dark Matter}

\author{M. R. Shahzad $^{a, b, c, d}$\orcidlink{0000-0001-9128-4018}}
\email{rizwanshahzad@zjut.edu.cn; dr.rizwanshahzad@bzu.edu.pk} 

\author{Hamza Rehman $^{a, b}$}
\email{hamzarehman244@zjut.edu.cn; hamzarehman244@gmail.com}

\author{Saddam Hussain$^{a, b}$\orcidlink{0000-0001-6173-6140}}
\email{saddamh@zjut.edu.cn}

\author{Tao Zhu $^{a, b}$\orcidlink{0000-0003-2286-9009}}
\email{Corresponding author: zhut05@zjut.edu.cn}

\affiliation{${}^{a}$ Institute for Theoretical Physics and Cosmology, Zhejiang University of Technology, Hangzhou 310023, China}
\affiliation{${}^{b}$ United Center for Gravitational Wave Physics (UCGWP), Zhejiang University of Technology, Hangzhou, 310023, China}
\affiliation{${}^{c}$ Department of Mathematics, Bahauddin Zakariya University, Vehari Campus, Vehari 61100, Pakistan}
\affiliation{${}^{d}$ Research Center of Astrophysics and Cosmology, Khazar University, Baku, AZ1096, 41 Mehseti Street, Azerbaijan.}

\date{\today}

\begin{abstract}

We investigate equatorial timelike motion of a massive test particle and the associated gravitational radiation around a Kerr black hole surrounded by perfect fluid dark matter. In the presence of a dark matter fluid, the radius of the innermost stable circular orbit decreases, while the specific angular momentum is modified and the specific energy increases as the dark matter (DM) parameter, $\alpha$, increases. We compute periodic timelike orbits using the rational frequency ratio $q$, with the $(z,w,v)$ classification, which is modified by the presence of DM while preserving the topological class. We analyze the gravitational waveforms corresponding to these periodic orbits and find that dark matter modifies the timing and distribution of the oscillatory features of the orbits. For a representative extreme-mass-ratio system, the corresponding characteristic strain is concentrated mainly in the millihertz (mHz) frequency band relevant to LISA. These results establish a direct connection between the dark-matter-induced modifications of strong-field orbital dynamics and the corresponding gravitational-wave signatures.
\end{abstract}

\maketitle
\section{Introduction}
Black holes provide natural laboratories for probing gravity in the strong-field regime. Observations of gravitational waves from compact-binary 
coalescences have opened a direct observational avenue for investigating the dynamics of black-hole systems and testing general relativity under extreme 
gravitational conditions~\cite{LIGOScientific:2016vbw,LIGOScientific:2016aoc, LIGOScientific:2016vlm,LIGOScientific:2016emj,LIGOScientific:2017bnn,
LIGOScientific:2017vox,LIGOScientific:2017ycc,LIGOScientific:2018mvr}. Future space-based observatories, in particular LISA, and Taiji, and Tianqin~\cite{LISA:2017pwj, Hu:2017mde, TianQin:2015yph, Gong:2021gvw, Bian:2025ifp}, are designed to probe the low-frequency gravitational-wave band, where extreme-mass-ratio inspirals (EMRIs) constitute an important class of sources~\cite{Hughes:2000ssa,Amaro-Seoane:2012lgq, Babak:2017tow,Glampedakis:2005hs}. In an EMRI, a stellar-mass compact object moves through the strong gravitational field of a massive black hole, and the resulting waveform carries detailed information about the geometry of the central spacetime~\cite{Hughes:2000ssa,Glampedakis:2005hs}. This sensitivity also makes EMRIs valuable probes of additional fields and possible departures 
from the standard Kerr description~\cite{Maselli:2021men,Liang:2022gdk}. Among the bound trajectories probing the strong-field region of an EMRI, particular attention is given to highly precessing orbits exhibiting zoom–whirl behavior. The dynamics of bound motion in the strong-field regime of black-hole
spacetimes exhibits a rich variety of precessing, periodic, separatrix, and zoom-whirl trajectories. Their orbital structure can be characterized through
rational relations between the fundamental frequencies, while the associated strong-field motion can leave distinctive features in the emitted
gravitational radiation~\cite{Levin:2008mq,Rana:2019bsn,Levin:2009sk,Healy:2009zm}. The connection between relativistic orbital motion and the corresponding
gravitational-wave signal can be studied within the numerical-kludge approach developed for EMRIs ~\cite{Babak:2006uv,poisson2014gravity}. These developments have stimulated a broad range of investigations of bound orbital dynamics and gravitational-wave signatures beyond the standard Kerr setting, including modified-gravity, quantum-corrected, hairy, and environmentally modified black-hole spacetimes
~\cite{Tu:2023xab,Li:2024tld,Haroon:2025rzx,Yang:2024lmj,Zhao:2024exh,Junior:2024tmi,Guo:2025scs,Choudhury:2025qsh,Chen:2025aqh,Lu:2025xlp, Ahmed:2025shr,Zhang:2025wni, Alloqulov:2025dqi,Sharipov:2025yfw, Ahmed:2025azu,Zare:2025aek, Wang:2025wob,Alloqulov:2025bxh,Lu:2025cxx,Wang:2025hla,Wang:2026ncv}.

Astrophysical black holes need not be regarded as isolated vacuum objects,  particularly when they reside at the centers of galaxies where the surrounding  matter distribution can influence the gravitational environment. In this context, 
it is natural to ask whether dark matter can leave discernible imprints on strong-field phenomena in the vicinity of a black hole. A useful phenomenological description is provided by the perfect-fluid dark-matter (PFDM) model, in which 
the dark-matter distribution is represented by an effective fluid source and is related to the approximately flat rotation curves observed on galactic scales~\cite{Rahaman:2010PFDM}. The corresponding rotating black-hole geometry 
was obtained in Ref.~\cite{Xu:2018PFDM}, providing a Kerr-like spacetime in which the effects of rotation and the surrounding dark-matter distribution can be examined within a common framework. The Kerr geometry is recovered when the PFDM contribution is removed. Since then, different strong-field aspects of this spacetime have been investigated. In particular, the PFDM parameter has been shown to modify the black-hole shadow~\cite{Hou:2018PFDMShadow}, characteristic 
orbital and precession frequencies~\cite{Rizwan:2019PFDMPrecession}, and the innermost stable circular orbit and radiative properties of thin accretion discs~\cite{HeydariFard:2023PFDMDisc}. These studies indicate that the surrounding dark-matter distribution can alter characteristic length and frequency scales of the strong-field geometry, although the magnitude of the effect depends on the observable and the parameter range considered. This motivates extending the analysis beyond photon propagation and circular-orbit observables to the bound motion of
massive particles, where the influence of the background geometry on the orbital dynamics can be examined directly.

The periodic timelike geodesics provide a systematic framework for characterizing strong-field bound motion. In the Kerr spacetime, bound equatorial motion can
be organized in terms of a discrete set of periodic trajectories. Levin and Perez-Giz showed that these orbits admit a rational-number classification
~\cite{Levin:2008mq}, which was subsequently developed into an energy-level description of black-hole orbits~\cite{Levin:2009sk}. In this framework, a periodic orbit is characterized by
\begin{equation}
q=\frac{\omega_{\phi}}{\omega_{r}}-1
  =w+\frac{v}{z},
\end{equation}
where $\omega_{\phi}$ and $\omega_{r}$ denote the azimuthal and radial frequencies, respectively, while the integers $(z,w,v)$ characterize the topology of the orbit. The strong-field members of this family naturally exhibit zoom-whirl behavior, with extended radial excursions interrupted by rapid azimuthal revolutions in the vicinity of the black hole. Such trajectories are of particular interest for extreme-mass-ratio systems, since the gravitational radiation generated by the small compact object retains detailed information about its orbital motion. 

The prospect of using periodic-orbit dynamics and their gravitational-wave signatures to probe the environment of a black hole has received increasing attention. Periodic orbits and their gravitational radiation have recently been examined for black holes embedded in different dark-matter distributions~\cite{Haroon:2025rzx,Alloqulov:2025Dehnen}. More directly, zoom-whirl dynamics and the associated gravitational waveforms have been studied for a charged black hole surrounded by PFDM~\cite{RawatNandan:2026PFDM}, while periodic orbits in an Euler-Heisenberg black hole with a PFDM environment have also been shown to carry characteristic modifications in their gravitational radiation~\cite{Gogoi:2026EHPFDM}. These developments demonstrate that the periodic-orbit sector is sensitive to environmental modifications of the background geometry and can therefore complement shadow, accretion, and circular-orbit observables. The existing analyses, however, are primarily based on static dark-matter geometries or on charged and nonlinear-electrodynamic PFDM black holes. To the best of our knowledge, the corresponding connection between rational periodic motion and gravitational-wave signatures has not yet been systematically developed for the uncharged rotating PFDM spacetime considered here, where frame dragging and the dark-matter contribution act simultaneously on the strong-field orbital dynamics.

The purpose of the present work is to investigate how the surrounding PFDM distribution modifies the strong-field bound dynamics of a rotating black hole and how these modifications are carried into the gravitational radiation from periodic motion. We first analyze equatorial timelike geodesics and determine the marginally bound orbit (MBO) and the innermost stable circular orbit (ISCO) for both prograde and retrograde motion, thereby establishing the range of strong-field bound dynamics relevant to the subsequent analysis. We then construct families of bound periodic orbits using the rational parameter $q=\omega_{\phi}/\omega_{r}-1$ and the associated $(z,w,v)$ classification, with particular attention to the dependence of their orbital energies and spatial structure on the PFDM parameter. The radiative imprint of these trajectories is subsequently studied within the numerical-kludge framework. We compute the two gravitational-wave polarizations in the time domain and evaluate the corresponding characteristic strains in order to determine how the PFDM-induced changes in the geodesic motion are transferred to the emitted gravitational-wave signal. For a representative extreme-mass-ratio system, the characteristic strain is also compared with the LISA sensitivity curve to place the resulting signals in the frequency range relevant to space-based gravitational-wave observations.

The remainder of the paper is organized as follows. In Sec. \ref{sec:theoretical_setup}, we introduce the rotating PFDM spacetime, derive the timelike geodesic equations for equatorial motion, and analyze the prograde and retrograde MBO and ISCO. Sec. \ref{sec:periodic_orbits}, is devoted to bound periodic motion, including the rational-orbit classification and the influence of the PFDM parameter on the orbital energy and geometry. In Sec. \ref{sec:GW_waveforms_setup}, we construct the gravitational waveforms associated with selected periodic trajectories and study their time-domain behavior and characteristic strains. Finally, Sec. \ref{sec:conclusion}, summarizes our main results and discusses their implications.


\section{Timelike geodesics and periodic motion}
\label{sec:theoretical_setup}
We consider the rotating black-hole spacetime surrounded by perfect fluid dark matter (PFDM), whose line element in Boyer-Lindquist coordinates $(t,r,\theta,\phi)$ is given by \cite{Xu:2018PFDM}
\begin{align}
ds^{2}={}& -\frac{\Delta-a^{2}\sin^{2}\theta}{\rho^{2}}\,dt^{2} +\frac{\rho^{2}}{\Delta}\,dr^{2} +\rho^{2}\,d\theta^{2} \nonumber\\ &-\frac{2a\sin^{2}\theta}{\rho^{2}} \left[ 2Mr-\alpha r\ln\left(\frac{r}{|\alpha|}\right) \right]dt\,d\phi \nonumber\\ &+\sin^{2}\theta \left[ r^{2}+a^{2} +\frac{a^{2}\sin^{2}\theta}{\rho^{2}} \left( 2Mr-\alpha r\ln\left(\frac{r}{|\alpha|}\right) \right) \right]d\phi^{2},
\label{metricPFDM}
\end{align}
where
\begin{equation}
\Delta = r^{2}+a^{2}-2Mr +\alpha r\ln\left(\frac{r}{|\alpha|}\right), \ \rho^{2}=r^{2}+a^{2}\cos^{2}\theta .
\label{metricfun} 
\end{equation}
Here, $M$ represents the black-hole mass, $a$ denotes the rotation parameter, and $\alpha$ determines the strength of the surrounding PFDM distribution. The logarithmic contribution proportional to $\alpha$ encodes the modification induced by the dark-matter environment and is associated with the approximately flat rotational velocity at large
galactic distances~\cite{Xu:2018PFDM}. In the limit $\alpha\rightarrow0$, the PFDM contribution disappears and the metric reduces to the Kerr geometry. For $a=0$, one instead recovers the corresponding static, spherically symmetric black-hole solution surrounded by PFDM,
\begin{equation}
\mathcal{F}(r)= 2Mr-\alpha r\ln\left(\frac{r}{|\alpha|}\right), \ \Delta=r^{2}+a^{2}-\mathcal{F}(r).
\label{Fdef}
\end{equation}
We now turn to the timelike geodesic motion of a massive test particle in the rotating PFDM spacetime. The corresponding Lagrangian can be written as
\begin{align}
2\mathcal{L}={}& -\left(1-\frac{\mathcal{F}(r)}{\rho^{2}}\right)\dot{t}^{\,2} -\frac{2a\mathcal{F}(r)\sin^{2}\theta}{\rho^{2}} \dot{t}\dot{\phi} +\frac{\rho^{2}}{\Delta}\dot{r}^{\,2} +\rho^{2}\dot{\theta}^{\,2}
\nonumber\\
&+\sin^{2}\theta \left[ r^{2}+a^{2} +\frac{a^{2}\mathcal{F}(r)\sin^{2}\theta}{\rho^{2}} \right]\dot{\phi}^{\,2}.
\label{lagrangian}
\end{align}
Here, an overdot denotes differentiation with respect to the particle's proper time $\tau$. For this Kerr-like geometry, the Hamilton-Jacobi equation remains separable, allowing the timelike geodesic equations to be expressed in the first-order form
\begin{align}
\rho^{2}\dot{t}
&= \frac{1}{\Delta} \left[ \mathcal{A}E-a\mathcal{F}(r)L \right],
\label{tdot}\\
\rho^{2}\dot{r} &= \pm\sqrt{\mathcal{R}(r)},
\label{rdot}\\
\rho^{2}\dot{\theta} &= \pm\sqrt{\Theta(\theta)},
\label{thetadot}\\
\rho^{2}\dot{\phi} &= \frac{1}{\Delta} \left[ a\mathcal{F}(r)E +\bigl(\rho^{2}-\mathcal{F}(r)\bigr) L\csc^{2}\theta \right].
\label{phidot}
\end{align}
Here,
\begin{equation}
\mathcal{A} = (r^{2}+a^{2})^{2} -a^{2}\Delta\sin^{2}\theta,
\label{Adef}
\end{equation}
while the radial and polar potentials are given by
\begin{equation}
\mathcal{R}(r) =\left[(r^{2}+a^{2})E-aL\right]^{2} -\Delta \left[ r^{2}+\mathcal{C}+(aE-L)^{2} \right],
\label{Rdef}
\end{equation}
and
\begin{equation}
\Theta(\theta) = \mathcal{C} + \left[ a^{2}(E^{2}-1)-L^{2}\csc^{2}\theta \right]\cos^{2}\theta .
\label{Thetadef}
\end{equation}
The quantities $E$ and $L$ denote, respectively, the conserved specific energy and specific azimuthal angular momentum of the particle, associated with the stationary and axisymmetric character of the spacetime. The additional constant $\mathcal{C}$ is the Carter constant arising from the separability of the Hamilton-Jacobi equation. Notably, the PFDM modification enters the radial potential through $\Delta$, while the angular potential $\Theta(\theta)$ retains its Kerr form.
\subsection{Particle motion on the equatorial plane}
We consider timelike particle motion on the equatorial plane, $\theta=\pi/2$, for which $\dot{\theta}=0$, $\mathcal{C}=0$, and $\rho^{2}=r^{2}$. The radial potential Eq. (\ref{Rdef}) reduces to

\begin{equation}
\mathcal{R}(r)= \left[(r^{2}+a^{2})E-aL\right]^{2} -\Delta\left[r^{2}+(aE-L)^{2}\right].
\label{ReqPFDM}
\end{equation}
The corresponding radial equation of motion can then be written as
\begin{eqnarray}
\dot{r}^{\,2} &=& E^{2} -\frac{r\Delta-r\left(a^{2}E^{2}-L^{2}\right)-\left[2M-\alpha\ln\left(\frac{r}{|\alpha|}\right)\right](aE-L)^{2}}{r^{3}}\nonumber\\ &=& E^{2}-V_{\rm eff}(r),
\label{radeqPFDM}
\end{eqnarray}
%
where the effective potential is given by
\begin{equation}
V_{\rm eff}(r) = \frac{r\Delta-r\left(a^{2}E^{2}-L^{2}\right)-\left[2M-\alpha\ln\left(\frac{r}{|\alpha|}\right)
\right](aE-L)^{2}}{r^{3}}.
\label{VeffPFDM}
\end{equation}

We now consider two characteristic circular orbits that play a central role in the dynamics of bound timelike particles: the marginally bound orbit (MBO) and the innermost stable circular orbit (ISCO). The MBO marks the threshold between bound and unbound motion and corresponds to an unstable circular orbit for which the radial velocity vanishes, $\dot r=0$, and the specific energy reaches the marginally bound value, $E_{\rm MBO}=1$. The ISCO, on the other hand, defines the innermost circular orbit that remains radially stable. Also, the MBO is computed by imposing the following condition 
\begin{equation}
V_{\rm eff}(r)=1, \qquad V_{\rm eff}'(r)=0,
\label{MBOcond}
\end{equation}
where a prime denotes differentiation with respect to $r$. Furthermore, the ISCO is obtained by imposing the following condition:
\begin{equation}
V_{\rm eff}(r)=E^{2},
V_{\rm eff}'(r)=0, V_{\rm eff}''(r)=0.
\label{ISCOcond}
\end{equation}

The resulting equations are solved numerically for prograde equatorial motion by setting $M=1$ and fixing $a/M=0.3$, while varying the PFDM parameter $\alpha/M$. The corresponding orbital quantities are summarized in Table~\ref{tab:isco_mbo_pro}. The numerical results show that the PFDM parameter systematically modifies the critical circular orbits. As $\alpha/M$ increases, both $r_{\rm ISCO}^{\rm P}$ and $r_{\rm MBO}^{\rm P}$ decrease, indicating that the marginally stable and marginally bound circular orbits shift inward toward the black hole. This inward displacement shows that the PFDM contribution modifies the strong-field orbital structure and allows both critical circular orbits to occur at smaller radii.

\begin{table}[t]
	\caption{ISCO and MBO parameters for retrograde equatorial timelike
		orbits around the rotating black hole surrounded by PFDM for different
		values of $\alpha/M$, with $a/M=0.3$. The marginally bound orbit
		satisfies $E_{\rm MBO}^{\rm P}, E_{\rm MBO}^{\rm R}=1$.}
	\label{tab:isco_mbo_retro}
	\label{tab:isco_mbo_pro}
	
	\begin{tabular}{|c|c|c|c|c|c|}
		\toprule
		\toprule
		$\alpha/M$ &
		$r_{\rm ISCO}^{\rm P}/M$ &
		$L_{\rm ISCO}^{\rm P}/M$ &
		$E_{\rm ISCO}^{\rm P}$ &
		$r_{\rm MBO}^{\rm P}/M$ &
		$L_{\rm MBO}^{\rm P}/M$ \\
		\bottomrule
		\toprule
		0.00 & 4.97862 & 3.15360 & 0.930642 & 3.37332 & 3.67332 \\
		\hline
		0.05 & 4.47708 & 2.81791 & 0.934244 & 3.04376 & 3.26011 \\
		\hline
		0.10 & 4.20601 & 2.61623 & 0.939040 & 2.86890 & 3.00141 \\
		\hline
		0.15 & 4.02850 & 2.46867 & 0.944534 & 2.75816 & 2.80506 \\
		\hline
		0.20 & 3.91312 & 2.35640 & 0.950508 & 2.69142 & 2.64929 \\
		\hline
		0.25 & 3.84418 & 2.26967 & 0.956793 & 2.65937 & 2.52275 \\
		\bottomrule
	\end{tabular}
	
	\begin{tabular}{|c|c|c|c|c|c|}
		\toprule
		$\alpha/M$ &
		$r_{\rm ISCO}^{\rm R}/M$ &
		$L_{\rm ISCO}^{\rm R}/M$ &
		$E_{\rm ISCO}^{\rm R}$ &
		$r_{\rm MBO}^{\rm R}/M$ &
		$L_{\rm MBO}^{\rm R}/M$ \\
		\bottomrule
		\toprule
		0.00 & 6.94927 & -3.72692 & 0.950824 & 4.58035 & -4.28035 \\
		\hline
		0.05 & 6.44197 & -3.36408 & 0.955273 & 4.25329 & -3.83530 \\
		\hline
		0.10 & 6.16721 & -3.13197 & 0.959925 & 4.08425 & -3.54078 \\
		\hline
		0.15 & 5.99021 & -2.95223 & 0.964771 & 3.98465 & -3.30620 \\
		\hline
		0.20 & 5.88120 & -2.80687 & 0.969744 & 3.93661 & -3.11064 \\
		\hline
		0.25 & 5.82627 & -2.68679 & 0.974780 & 3.93395 & -2.94329 \\
		\bottomrule
		\bottomrule
	\end{tabular}
\end{table}

The specific angular momentum exhibits a similar monotonic trend, with both $L_{\rm ISCO}^{\rm P}$ and $L_{\rm MBO}^{\rm P}$ decreasing as the PFDM parameter increases. Thus, the marginally stable and marginally bound circular orbits are associated with progressively smaller values of the specific angular momentum. The behavior of the specific energy is qualitatively different, with $E_{\rm ISCO}^{\rm P}$ increasing monotonically by approximately $3.5\%$ over the considered range of $\alpha/M$, approaching the marginally bound value $E=1$. In the present analysis, $E=1$ corresponds to the marginally bound limit, and the quantity $1-E_{\rm ISCO}^{\rm P}$ represents the specific binding energy of the ISCO relative to infinity. The increase of $E_{\rm ISCO}^{\rm P}$ with $\alpha/M$ therefore implies a decrease in the binding energy. Thus, although $r_{\rm ISCO}^{\rm P}$ shifts inward as the PFDM parameter increases, the corresponding ISCO becomes less tightly bound. This shows that, in the PFDM-modified geometry, the radial location of the ISCO and its binding energy need not exhibit the same trend.

Taken together, these trends show that the PFDM contribution modifies the locations, angular momenta, and binding properties of the critical prograde equatorial circular orbits. In the absence of PFDM ($\alpha=0$), the results reduce to the standard prograde Kerr values for $a/M=0.3$. Furthermore, taking the simultaneous limits $\alpha\rightarrow0$ and $a\rightarrow0$ recovers the corresponding Schwarzschild results.

We next consider the retrograde equatorial branch, for which the specific angular momentum is negative for $a>0$. The corresponding ISCO and MBO parameters are presented in Table~\ref{tab:isco_mbo_retro}. For a fixed value of $\alpha/M$, the retrograde critical orbits are located at larger radii than their prograde counterparts, reflecting the prograde-retrograde asymmetry associated with the black-hole rotation. As $\alpha/M$ increases, both $r_{\rm ISCO}^{\rm R}$ and $r_{\rm MBO}^{\rm R}$ decrease monotonically. In particular, $r_{\rm ISCO}^{\rm R}$ decreases by approximately $16.2\%$, while $r_{\rm MBO}^{\rm R}$ decreases by approximately $14.1\%$ over the parameter range considered. Thus, the PFDM contribution shifts both retrograde critical circular orbits inward over the parameter range considered. The specific angular momenta $L_{\rm ISCO}^{\rm R}$ and $L_{\rm MBO}^{\rm R}$ exhibit a monotonic increase with increasing $\alpha/M$. In terms of their magnitudes, $|L_{\rm ISCO}^{\rm R}|$ and $|L_{\rm MBO}^{\rm R}|$ decrease by approximately $27.9\%$ and $31.2\%$, respectively. Thus, although the angular momenta remain negative throughout the retrograde branch, their values move progressively toward zero as the PFDM parameter increases. The specific energy of the retrograde ISCO also increases monotonically, by approximately $2.5\%$ over the considered range of $\alpha/M$. Since $E=1$ corresponds to the marginally bound limit, this increase implies a decrease in the specific binding energy of the ISCO. Hence, although the retrograde ISCO moves inward as the PFDM parameter increases, it becomes progressively less tightly bound.

A comparison of the two orbital branches shows that, for the same value of $\alpha/M$, the retrograde ISCO and MBO remain farther from the black hole than their prograde counterparts. The prograde specific angular momenta decrease with increasing $\alpha/M$, whereas the retrograde angular momenta increase from negative values toward zero. The PFDM contribution therefore modifies the two orbital branches differently in angular momentum while preserving the characteristic prograde-retrograde asymmetry induced by rotation.

\section{Periodic orbits}
\label{sec:periodic_orbits}

We now investigate periodic bound timelike orbits in the equatorial plane of the rotating PFDM black hole. In contrast to circular motion, a bound eccentric orbit undergoes radial oscillations while simultaneously advancing in the azimuthal direction. The radial motion is confined between the periapsis $r_{\rm p}$ and the apoapsis
$r_{\rm a}$, corresponding to the inner and outer turning points, respectively. At these points, $\dot r=0$, and therefore
\begin{equation}
\mathcal{R}(r_{\rm p})=\mathcal{R}(r_{\rm a})=0, \qquad r_{\rm p}<r_{\rm a}.
\label{turningpoints}
\end{equation}

During one complete radial period, the particle moves from $r_{\rm p}$ to $r_{\rm a}$ and returns to $r_{\rm p}$ while accumulating an azimuthal angle $\Delta\phi$. The corresponding azimuthal advance is
\begin{equation}
\Delta\phi =2\int_{r_{\rm p}}^{r_{\rm a}} \frac{d\phi}{dr}\,dr =2\int_{r_{\rm p}}^{r_{\rm a}}\frac{\dot{\phi}}{\dot r}\,dr ,
\label{Deltaphi}
\end{equation}
where the integral is evaluated along the outward radial branch, and the factor of $2$ accounts for the subsequent return from $r_{\rm a}$ to $r_{\rm p}$. To classify the resulting closed trajectories, we adopt the periodic-orbit taxonomy introduced by Levin and Perez-Giz~\cite{Levin:2008mq} and subsequently developed and
applied to relativistic black-hole orbits in Refs.~\cite{Levin:2009sk,Misra:2010pu,Babar:2017gsg, Levin:2008yp}.
The radial and azimuthal motions are characterized by the rational parameter
\begin{equation}
q\equiv \frac{\omega_{\phi}}{\omega_{r}}-1=\frac{\Delta\phi}{2\pi}-1=w+\frac{v}{z},
\label{qdef}
\end{equation}
where $\omega_{r}$ and $\omega_{\phi}$ denote the radial and azimuthal frequencies, respectively. Equivalently, using Eq.~(\ref{Deltaphi}),
\begin{equation}
q=\frac{1}{\pi}\int_{r_{\rm p}}^{r_{\rm a}}\frac{\dot{\phi}}{\dot r}\,dr-1 .
\label{qintegral}
\end{equation}
This form is particularly convenient for determining $q$ directly from the radial turning points of a bound orbit. The quantity $q$ measures the relative advance of the azimuthal motion with respect to one radial oscillation. For rational $q$, the radial and azimuthal motions repeat in a fixed rational ratio, and the trajectory closes after a finite number of radial periods. For irrational $q$, no finite repetition occurs and the orbit continues to precess around the black hole. The integers $(z,w,v)$ specify the topology of a periodic orbit. The integer $z$ gives the number of distinct leaves traced before the orbit closes, $w$ denotes the number of whirls near periapsis during each radial cycle, and $v$ determines the order in which the successive leaves are visited~\cite{Levin:2008mq}. For $z>1$, the integer $v$ satisfies
\begin{equation}
1\leq v\leq z-1,
\end{equation}
with $v$ and $z$ relatively prime, whereas a single-leaf orbit $(z=1)$ is assigned $v=0$. This restriction removes degenerate representations and provides a unique topological label for each closed orbit. In the numerical analysis, we fix the specific angular momentum $L$ and vary the energy $E$ to obtain prescribed rational values of $q$. For each value of the PFDM parameter $\alpha/M$, the radial equation is required to admit two physical turning points outside the event horizon. The corresponding value of $E$ is then used to integrate the geodesic equations and construct the associated periodic trajectory. The dependence of the rational parameter on the conserved orbital quantities has been investigated in related black-hole spacetimes~\cite{Azreg-Ainou:2020bfl}.

\begin{table*}[!t]
\centering

\captionsetup{margin={0cm,0.3cm}}
\caption{Specific energies of selected prograde periodic orbits for different
values of the PFDM parameter $\alpha/M$, with $M=1$, $a/M=0.3$, and
$L/M=2.8$. The orbital configurations are labeled by the triplet $(z,w,v)$,
with $q=w+v/z$.}
\label{tab:energy_periodic}

\setlength{\tabcolsep}{5pt}
\renewcommand{\arraystretch}{1.15}

\resizebox{0.95\textwidth}{!}{%
\begin{tabular}{|c|c|c|c|c|c|c|c|}
\toprule
\toprule
$\alpha/M$ &
$L/M$ &
$E_{(1,1,0)}$ &
$E_{(4,1,1)}$ &
$E_{(3,1,1)}$ &
$E_{(2,1,1)}$ &
$E_{(5,1,3)}$ &
$E_{(1,2,0)}$ \\
\bottomrule
\toprule
0.10 & 2.8 & 0.95993187 & 0.96309104 & 0.96363900 &
0.96434282 & 0.96460212 & 0.96506552 \\
\hline
0.11 & 2.8 & 0.96701449 & 0.97009043 & 0.97061550 &
0.97128199 & 0.97152393 & 0.97194602 \\
\hline
0.12 & 2.8 & 0.97401056 & 0.97702674 & 0.97753462 &
0.97817273 & 0.97840141 & 0.97879217 \\
\hline
0.13 & 2.8 & 0.98089993 & 0.98388199 & 0.98437863 &
0.98499726 & 0.98521652 & 0.98558451 \\
\hline
0.14 & 2.8 & 0.98765245 & 0.99063353 & 0.99112685 &
0.99173755 & 0.99195213 & 0.99230693 \\
\bottomrule
\toprule
\end{tabular}%
}

\end{table*}
\begin{figure*}[t]
\centering
\includegraphics[scale=0.55]{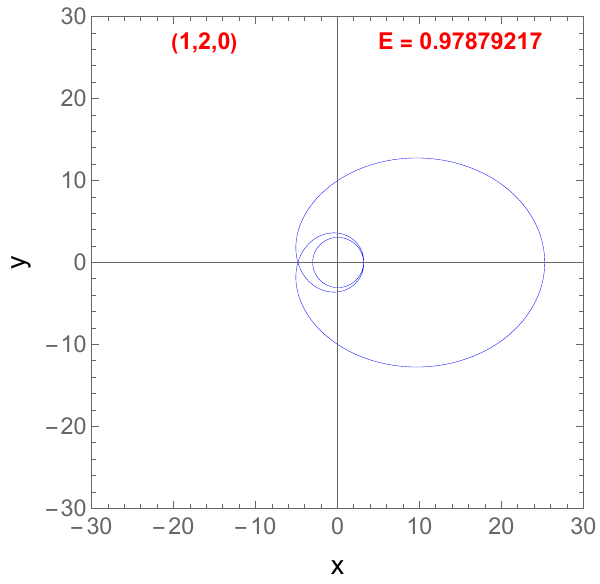}
\includegraphics[scale=0.55]{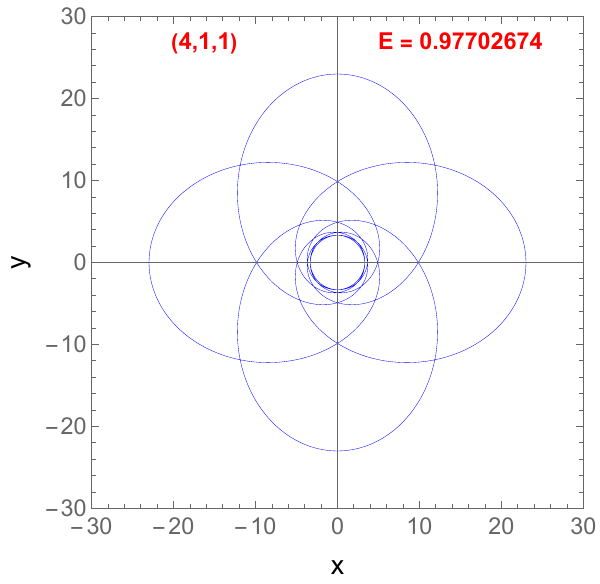}
\includegraphics[scale=0.55]{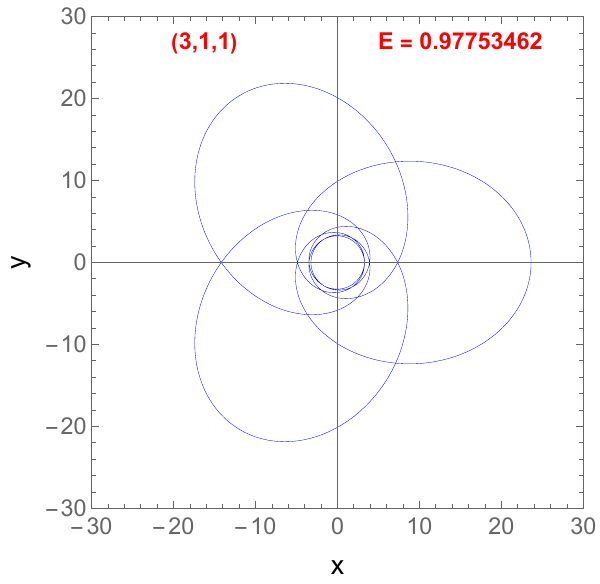}
\includegraphics[scale=0.55]{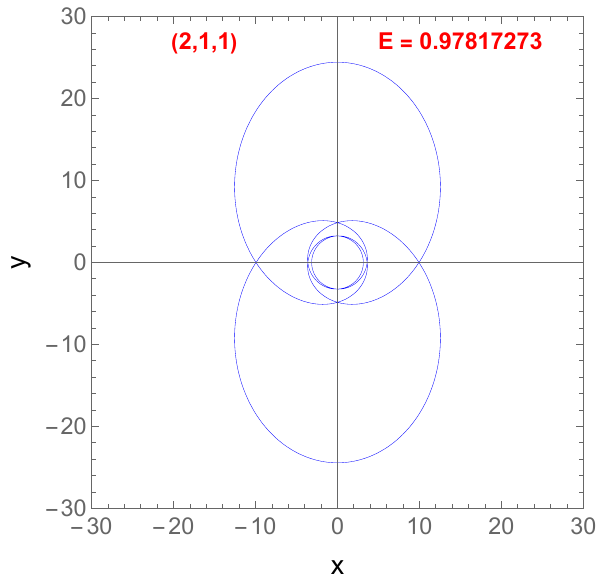}
\includegraphics[scale=0.55]{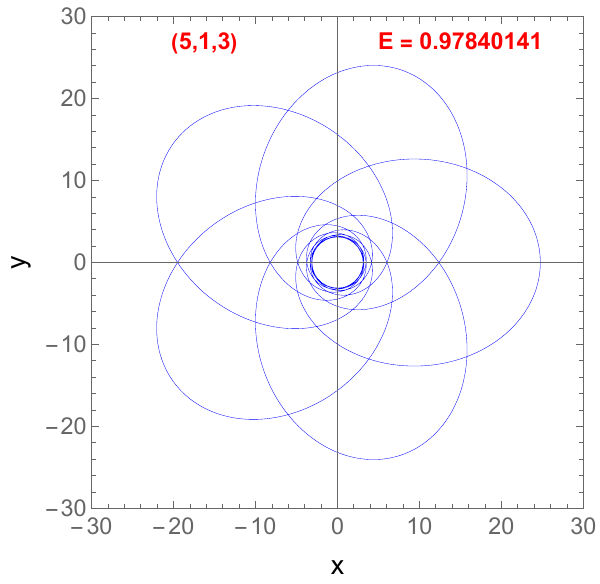}
\includegraphics[scale=0.55]{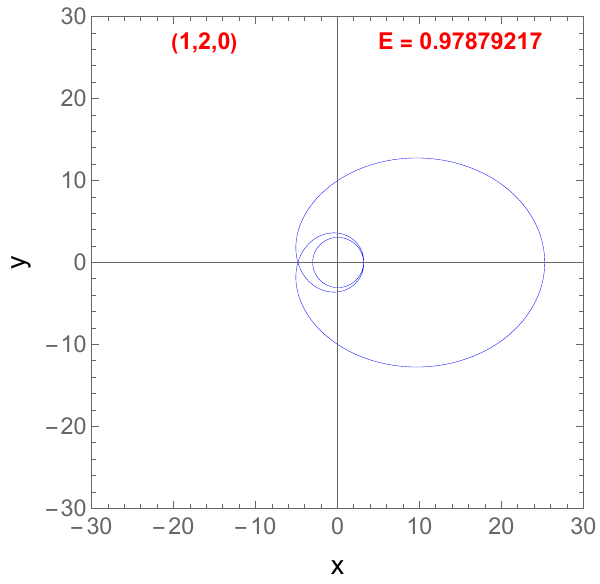}
\captionsetup{margin={0cm,1.5cm}}
\caption{Representative prograde periodic orbits for $\alpha/M=0.12$, with $M=1$, $a/M=0.3$, and $L/M=2.8$. The orbital configurations are labeled by the triplet $(z,w,v)$, corresponding to the rational parameter $q=w+v/z$.}
\label{fig:periodic_orbits}
\end{figure*}
Figure~\ref{fig:periodic_orbits} illustrates representative prograde periodic orbits for $\alpha/M=0.12$, with $M=1.0$, $a/M=0.30$, and $L/M=2.80$. The different orbital configurations are classified by the triplet $(z,w,v)$, or equivalently by the rational parameter $q=w+v/z$. The cases $(1,1,0)$ and $(1,2,0)$ correspond to single-leaf orbits with $q=1$ and $q=2$, respectively, while the configurations $(4,1,1)$, $(3,1,1)$, $(2,1,1)$, and $(5,1,3)$ exhibit four, three, two, and five distinct leaves. The integer $w$ controls the number of whirls near periapsis, whereas $v$ determines the order in which the successive leaves are visited. For fixed $\alpha/M=0.12$ and $L/M=2.8$, the orbital energy increases with $q$, from $E=0.97401056$ for $q=1$ to $E=0.97879217$ for $q=2$. The corresponding intermediate configurations occur at energies lying between these two values. Thus, different rational values of $q$ select distinct closed trajectories with different topologies while remaining within a relatively narrow energy interval.

The corresponding specific energies are listed in Table~\ref{tab:energy_periodic}. Over the parameter range considered, $E$ increases systematically with $\alpha/M$ for each periodic-orbit family. At fixed $\alpha/M$, the energy also increases with $q$, showing that periodic trajectories with larger azimuthal advance per radial cycle occur at higher specific energies.

\begin{figure*}
\centering
\includegraphics[scale=0.55]{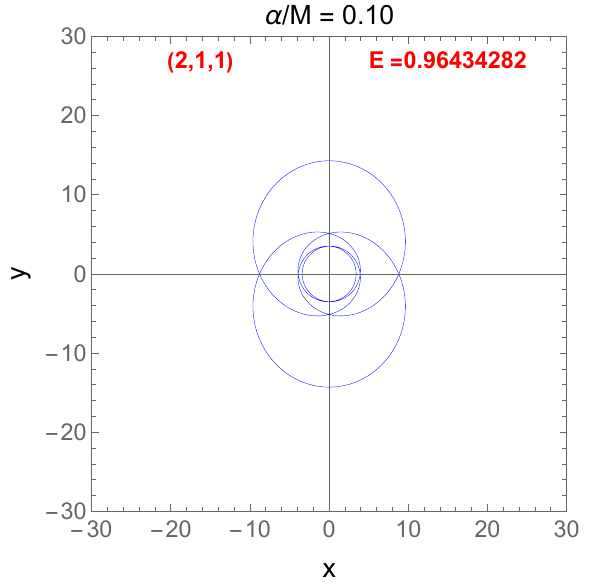}
\includegraphics[scale=0.55]{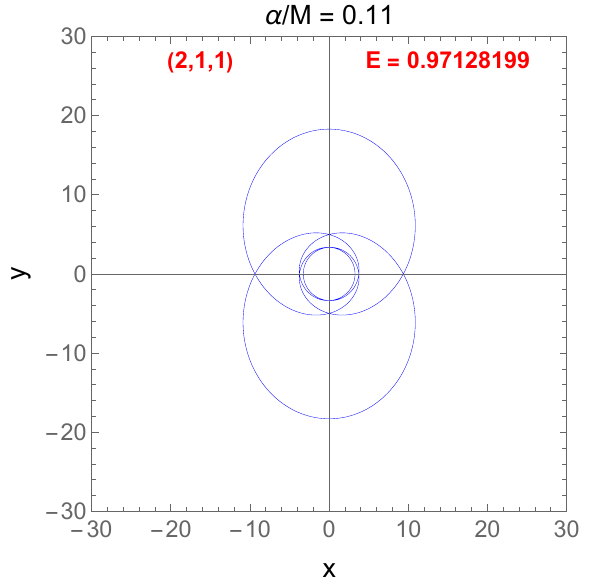}
\includegraphics[scale=0.55]{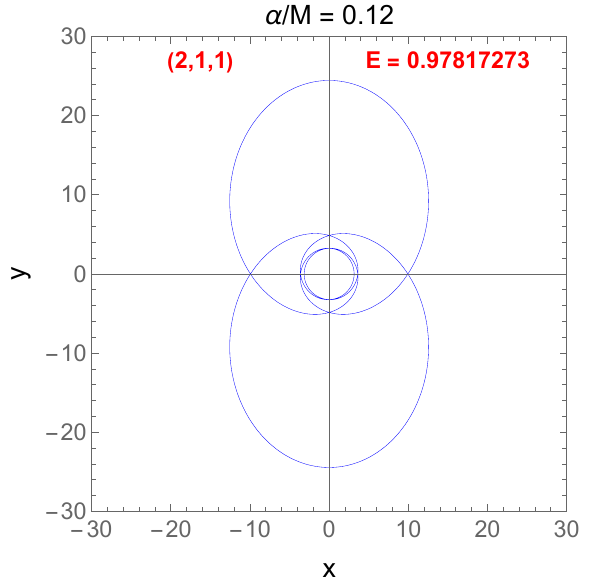}
\includegraphics[scale=0.55]{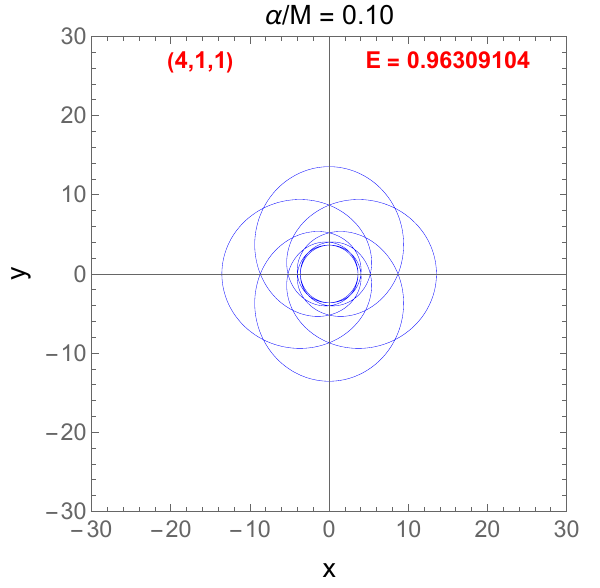}
\includegraphics[scale=0.55]{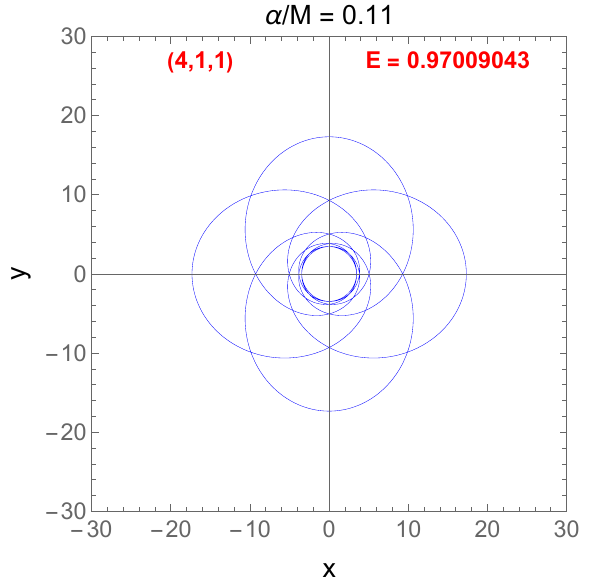}
\includegraphics[scale=0.55]{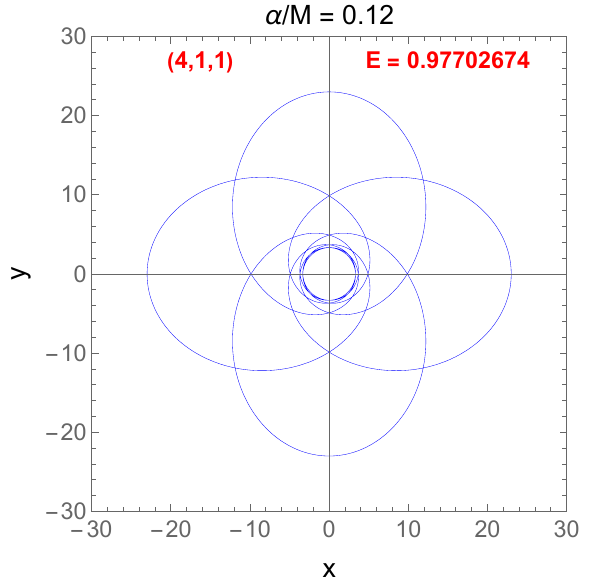}
\captionsetup{margin={0cm,1.5cm}}
\caption{Representative prograde periodic orbits for $(z,w,v)=(2,1,1)$, corresponding to $q=3/2$ (top row), and $(z,w,v)=(4,1,1)$, corresponding to $q=5/4$ (bottom row), for $\alpha/M=0.10$, $0.11$, and $0.12$. The spin and specific angular momentum are fixed at $a/M=0.3$ and $L/M=2.8$, respectively.}
\label{fig:alpha_orbits}
\end{figure*}

Figure~\ref{fig:alpha_orbits} illustrates the effect of the PFDM parameter on two representative prograde periodic-orbit families, $(z,w,v)=(2,1,1)$ and $(4,1,1)$, for $M=1$, $a/M=0.3$, and
$L/M=2.8$. The upper panels correspond to $q=3/2$ and exhibit two distinct leaves, whereas the lower panels correspond to $q=5/4$ and contain four leaves. In both cases, $w=1$, so that the trajectories
possess the same whirl number, while the difference in $z$ determines the number of leaves traced before the orbit closes. For a fixed orbital topology, increasing $\alpha/M$ from $0.10$ to $0.12$ produces a visible change in the radial extent of the trajectory, with the outer portions of the orbit extending to larger radii. The corresponding orbital energy also increases. For the $(2,1,1)$ family, the energy rises from $E=0.96434282$ at $\alpha/M=0.10$ to $E=0.97817273$ at $\alpha/M=0.12$, while for the
$(4,1,1)$ family it increases from $E=0.96309104$ to $E=0.97702674$ over the same interval. Thus, although the topological classification $(z,w,v)$ is preserved, the PFDM parameter modifies both the energy required to realize a given periodic orbit and its spatial structure. At fixed $\alpha/M$, the $(2,1,1)$ orbit requires a slightly larger energy than the $(4,1,1)$ orbit, consistent with the values listed in Table~\ref{tab:energy_periodic}. These results show that the PFDM
environment changes the quantitative properties of the periodic trajectories without altering their prescribed topological class.
\section{Gravitational-wave signatures of periodic orbits}
\label{sec:GW_waveforms_setup}
This section focuses on the gravitational radiation emitted by a particle moving on periodic timelike orbits around a rotating black hole surrounded by perfect fluid dark matter (PFDM). Such orbital
configurations are relevant to extreme-mass-ratio inspirals (EMRIs), in which a stellar-mass compact object moves in the strong gravitational field of a supermassive black hole. These systems are
among the most promising sources for future space-based gravitational-wave detectors, including LISA, Taiji, and TianQin.~\cite{Hu:2017mde,Gong:2021gvw,LISA:2017pwj}. Gravitational waves from EMRIs probe the strong-field orbital dynamics and are sensitive to the spacetime geometry of the central black hole.
In the present model, the PFDM contribution modifies the geodesic motion of the compact object and hence the structure of the periodic orbits. These modifications can propagate into the emitted waveform,
providing a possible means of identifying the influence of the surrounding PFDM through its gravitational-wave signatures. The gravitational waveforms from EMRIs are studied within the adiabatic approximation, which is valid when the inspiral timescale is much longer than the orbital period~\cite{hughes2000evolution,Barack:2003fp}.
In this regime, the motion of the compact object can be described by a sequence of geodesics of the rotating PFDM spacetime, since its energy and angular momentum evolve only slowly. For the short-term orbital dynamics considered here~\cite{Isoyama:2021jjd}, the radiation reaction, namely the backreaction of the emitted gravitational waves on the particle motion, is neglected at the leading order.

To construct the gravitational waveforms associated with the periodic orbits, we adopt the numerical-kludge prescription \cite{Babak:2006uv}. The procedure is implemented in two steps. First, the trajectory of the compact object is obtained by numerically integrating the timelike geodesic equations of the rotating PFDM spacetime. The resulting relativistic orbit is then used as the source of gravitational radiation, which is evaluated at leading quadrupole order. In this way, the strong-field orbital dynamics determined by the spin $a$ and the PFDM parameter $\alpha$ are retained, while the wave generation is treated within the standard quadrupole approximation. This semi-relativistic approach has been widely used to model gravitational-wave signals from EMRIs \cite{Barack:2003fp, Gair:2004iv, Hughes:2000ssa}. In terms of the mass quadrupole moment $I_{ij}$, the metric perturbation $h_{ij}$ in the radiation zone is given by

\begin{equation}
h_{ij}=\frac{1}{\mathcal{N}}\,\ddot{I}_{ij},
\label{eq:quadrupoleGW}
\end{equation}
where $\mathcal{N}=\frac{c^{4}D_L}{2G}$, $G=c=1$ with $D_L$ indicate the luminosity distance to the source. The trajectory $Z^{i}(t)$ of the compact object is obtained by numerically integrating the timelike geodesic equations of the rotating PFDM spacetime and is subsequently used to construct the
gravitational-wave signal. For a compact object of mass $\mu$ moving along $Z^{i}(t)$, the mass quadrupole moment is written as\cite{Thorne:1980ru}

\begin{equation}
I^{ij}=\mu \int d^{3}x\,x^{i}x^{j}\,\delta^{(3)}\!\left[x^{i}-Z^{i}(t)\right].
\label{eq:quadrupole}
\end{equation}
The geodesic equations are usually solved in Boyer-Lindquist coordinates $(r,\theta,\phi)$, whereas the quadrupole radiation formula is most conveniently evaluated in a Cartesian representation
of the source trajectory. Following the numerical-kludge prescription, we introduce the auxiliary Cartesian coordinates
\begin{equation}
x=r\sin\theta\cos\phi,\qquad
y=r\sin\theta\sin\phi,\qquad
z=r\cos\theta .
\label{eq:Cartesian}
\end{equation}
For the equatorial periodic orbits considered here, $\theta=\pi/2$, so that
\begin{equation}
x=r\cos\phi,\qquad
y=r\sin\phi,\qquad
z=0.
\label{eq:Cartesian_equatorial}
\end{equation}
The functions $r(t)$ and $\phi(t)$ obtained from the timelike geodesic equations of the rotating PFDM spacetime therefore determine the Cartesian trajectory used in the quadrupole calculation. The
corresponding position, velocity, and acceleration are then employed to construct the metric perturbation $h_{ij}$ and, subsequently, the two gravitational-wave polarizations.
The Cartesian representation of the orbit is used to construct the mass quadrupole moment $I_{ij}$, from which the gravitational-wave metric perturbation is obtained. The corresponding metric perturbation is obtained from the second time derivative of $I_{ij}$. For a compact object of mass
$\mu$, one finds
\begin{equation}
h_{ij} = \frac{\mu}{\mathcal{N}} \left(a_i x_j+a_j x_i+2v_i v_j\right),
\label{eq:hijCartesian}
\end{equation}
where $v_i=dx_i/dt$ and $a_i=d^2x_i/dt^2$ are the Cartesian velocity and acceleration components of the compact object, respectively, and $\mathcal{N}=c^4 D_L/(2G)$. The spin parameter $a$ and the PFDM parameter $\alpha$ enter the waveform through their influence on the geodesic trajectory and, consequently, on the Cartesian position, velocity, and acceleration of the compact object. This construction follows the numerical-kludge prescription commonly employed in modeling gravitational-wave signals from EMRIs
\cite{Gair:2004iv, Barack:2003fp,Babak:2006uv,Hughes:2000ssa}.

To describe the gravitational-wave signal measured by a distant observer, we introduce an observer-adapted Cartesian frame $(X,Y,Z)$ centered on the black hole. The direction of the observer with respect to the source frame $(x,y,z)$ is specified by the inclination angle $\iota$ and the azimuthal angle $\zeta$ ~\cite{Barack:2003fp,Babak:2006uv,Hughes:2000ssa}. The metric perturbation $h_{ij}$ obtained in the source frame is then projected onto the transverse plane of the observer frame to determine the two gravitational-wave polarizations. The corresponding orthonormal basis vectors, expressed in the $(x,y,z)$ coordinates, are

\begin{align}
\hat{\mathbf e}_{X}
&=
(\cos\zeta,-\sin\zeta,0),
\label{eq:eX}\\
\hat{\mathbf e}_{Y}
&=
(\cos\iota\sin\zeta,
 \cos\iota\cos\zeta,
 -\sin\iota),
\label{eq:eY}\\
\hat{\mathbf e}_{Z}
&=
(\sin\iota\sin\zeta,
 \sin\iota\cos\zeta,
 \cos\iota).
\label{eq:eZ}
\end{align}
The two gravitational-wave polarizations, $h_{+}$ and $h_{\times}$, are obtained by projecting the metric perturbation $h_{ij}$ given in Eq.~(\ref{eq:hijCartesian}) onto the detector-adapted frame,
\begin{align}
h_{+}&=\frac{1}{2}
\left(e_X^{i}e_X^{j}-e_Y^{i}e_Y^{j}\right)h_{ij},
\label{eq:hplus}\\
h_{\times}&=\frac{1}{2}\left(e_X^{i}e_Y^{j}+e_Y^{i}e_X^{j}\right)h_{ij}.
\label{eq:hcross}
\end{align}
The two polarization modes may also be written in terms of the
projected components of $h_{ij}$ in the observer-adapted frame.
Introducing $h_{\zeta\zeta}$, $h_{\iota\iota}$, and
$h_{\iota\zeta}$, we have
\begin{equation}
h_{+}
=
\frac{1}{2}
\left(
h_{\zeta\zeta}-h_{\iota\iota}
\right),
\qquad
h_{\times}=h_{\iota\zeta}.
\label{eq:polarizations}
\end{equation}

Also, the components determined as \cite{Babak:2006uv}
\begin{eqnarray}
h_{\zeta\zeta}&=& h_{xx}\cos^{2}\zeta-h_{xy}\sin 2\zeta+h_{yy}\sin^{2}\zeta ,
\label{eq:hzetazeta}
\\
h_{\iota\iota}&=&\cos^{2}\iota\left(h_{xx}\sin^{2}\zeta+h_{xy}\sin 2\zeta+h_{yy}\cos^{2}\zeta\right)
\nonumber\\&&+h_{zz}\sin^{2}\iota-\sin 2\iota\left(h_{xz}\sin\zeta+h_{yz}\cos\zeta\right),
\label{eq:hiotaiota}
\\h_{\iota\zeta}&=&\cos\iota\left[\frac{1}{2}h_{xx}\sin 2\zeta+h_{xy}\cos 2\zeta-\frac{1}{2}h_{yy}\sin 2\zeta\right]\nonumber\\&&+\sin\iota\left(h_{yz}\sin\zeta-h_{xz}\cos\zeta\right).
\label{eq:hiotazeta}
\end{eqnarray}

For the waveform calculations, we consider a compact object of mass $\mu=10M_{\odot}$ orbiting a supermassive black hole of mass $M=10^{7}M_{\odot}$. We set $\iota=\zeta=\frac{\pi}{4}$, $D_L=2~{\rm Gpc}$, and fix the spin and specific angular momentum at $a/M=0.3$ and $L/M=2.8$, respectively. The PFDM parameter $\alpha/M$ is varied, while for each periodic-orbit family the corresponding specific energy is chosen from the values obtained in the preceding section.


\begin{figure*}[t]
\centering

\begin{minipage}[c]{0.54\textwidth}
    \centering
    \includegraphics[width=\linewidth]{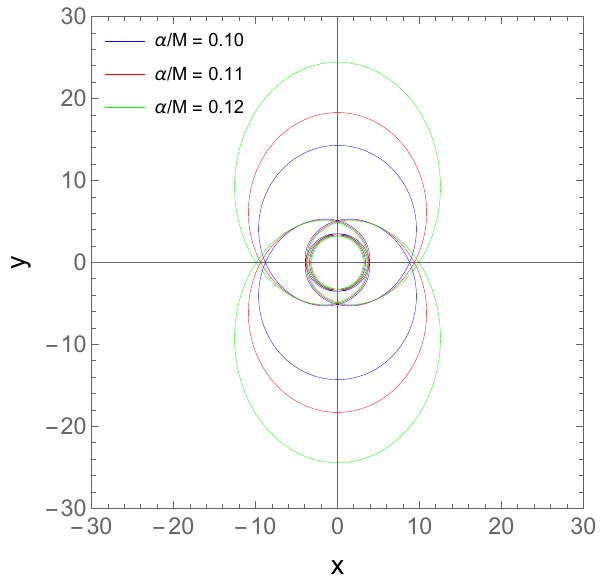}
\end{minipage}%
\hfill
\begin{minipage}[c]{0.46\textwidth}
    \centering
    \includegraphics[width=\linewidth]{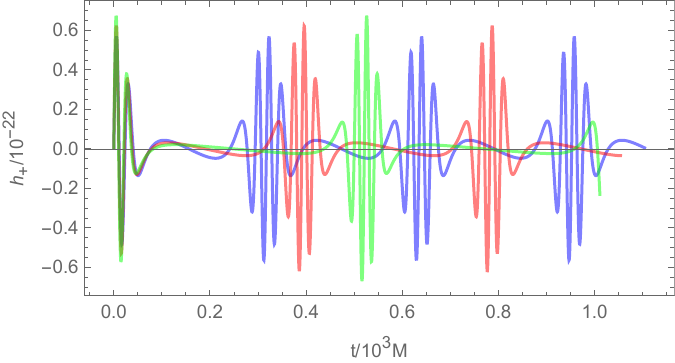}

    \vspace{-1mm}

    \includegraphics[width=\linewidth]{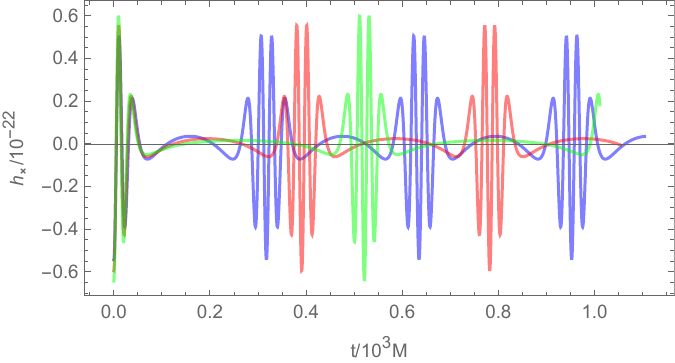}
\end{minipage}
\captionsetup{
    justification=justified,
    singlelinecheck=false,
    margin={0.1cm,0cm}
}
\caption{The $q=3/2$ periodic orbit, corresponding to $(z,w,v)=(2,1,1)$,
and the associated gravitational-wave polarizations $h_{+}$ and $h_{\times}$
for $\alpha/M=0.10$ (blue), $0.11$ (red), and $0.12$ (green). The spin and
specific angular momentum are fixed at $a/M=0.3$ and $L/M=2.8$, respectively.
The left panel shows the corresponding orbital trajectories, while the upper
and lower right panels display the plus and cross polarizations, respectively.}
\label{fig:GW-q32}
\end{figure*}

\begin{figure*}[t]
\centering

\begin{minipage}[c]{0.54\textwidth}
    \centering
    \includegraphics[width=\linewidth]{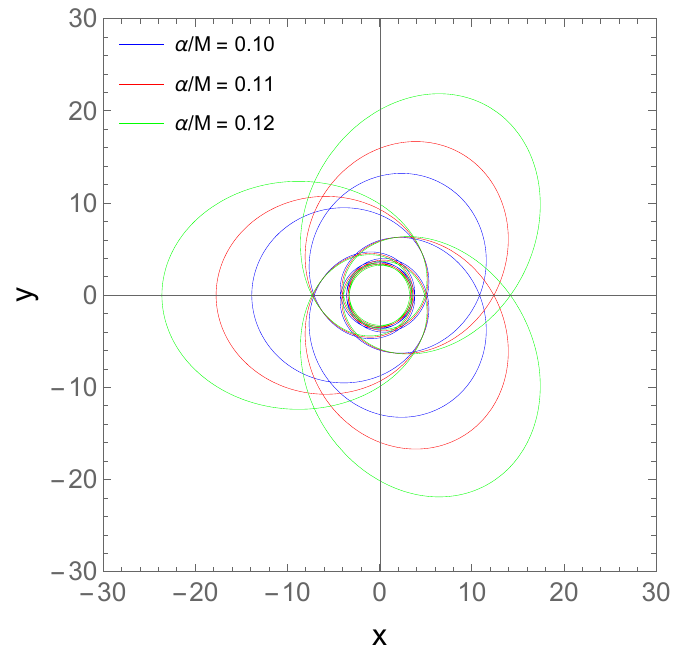}
\end{minipage}%
\hfill
\begin{minipage}[c]{0.46\textwidth}
    \centering
    \includegraphics[width=\linewidth]{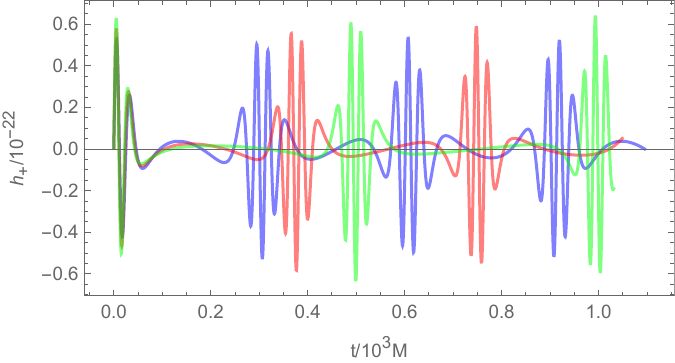}

    \vspace{-1mm}

    \includegraphics[width=\linewidth]{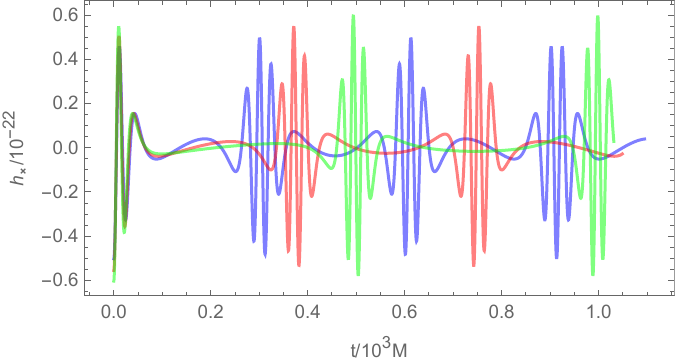}
\end{minipage}
\caption{The $q=4/3$ periodic orbit, corresponding to $(z,w,v)=(3,1,1)$, and the associated gravitational-wave polarizations $h_{+}$ and $h_{\times}$ for $\alpha/M=0.10$ (blue), $0.11$ (red), and $0.12$ (green). The spin and specific angular momentum are fixed at $a/M=0.3$ and $L/M=2.8$, respectively. The left panel shows the periodic trajectories, while the upper and lower right panels display $h_{+}$ and $h_{\times}$, respectively.}
\label{fig:GW-q43}
\end{figure*}
Figures~\ref{fig:GW-q32} and~\ref{fig:GW-q43} show the periodic trajectories and the corresponding gravitational-wave polarizations for the $q=3/2$ and $q=4/3$ orbital families, respectively. In both cases, increasing $\alpha/M$ from $0.10$ to $0.12$ leads to an increase in the radial extent of the orbit, while the characteristic topology of each periodic configuration is preserved. The $(2,1,1)$ orbit retains its two-leaf structure, whereas the $(3,1,1)$ orbit maintains the corresponding three-leaf pattern. This shows that, for the parameter range considered, the PFDM contribution mainly changes the radial scale of the periodic motion without altering its prescribed rational topology. The associated $h_{+}$ and $h_{\times}$ waveforms exhibit a sequence of relatively weak oscillations interrupted by short intervals of rapidly varying, larger-amplitude radiation. This structure is directly associated with the zoom-whirl dynamics of the underlying trajectories: the smoother parts of the signal are associated with motion at larger radial distances, whereas the more pronounced oscillations occur when the compact object approaches the strong-field region and undergoes rapid azimuthal motion. The different orbital topologies consequently produce distinct temporal patterns in both polarization modes. A variation of $\alpha/M$ also changes the location and separation of the oscillatory features along the time axis. These changes arise from the influence of the PFDM parameter on the underlying geodesic motion, which modifies the radial and azimuthal evolution of the compact
object and, consequently, the position, velocity, and acceleration entering the quadrupole waveform. In contrast, the peak amplitudes do not exhibit the monotonic dependence on $\alpha/M$. The influence of the PFDM parameter is instead more clearly seen in the change in the radial extent of the orbit and in the timing and distribution of the oscillations in $h_{+}(t)$ and $h_{\times}(t)$.
The gravitational-wave signals generated by the periodic orbits can also be characterized in the frequency domain through the characteristic strain,
\begin{equation}
h_c(f)=2f \left[|\widetilde h_{+}(f)|^2+|\widetilde h_{\times}(f)|^2\right]^{1/2}.
\label{eq:characteristic_strain}
\end{equation}
The Fourier amplitudes entering Eq.~\eqref{eq:characteristic_strain} are obtained by applying a discrete Fourier transform (DFT) to the time-domain polarizations $h_{+}(t)$ and $h_{\times}(t)$.
\begin{figure*}[ht!]
\centering
\includegraphics[width=0.49\textwidth]{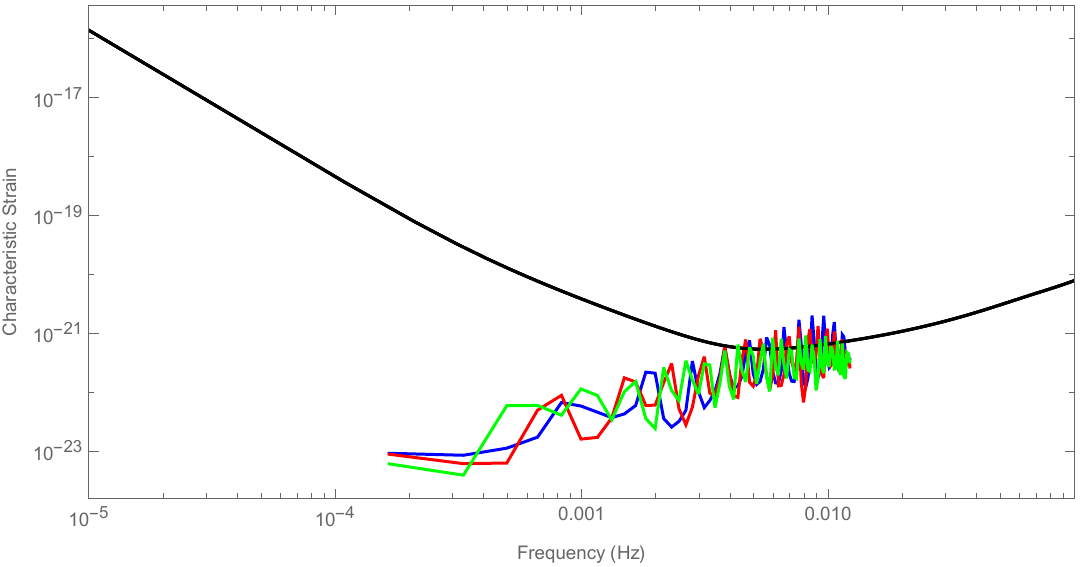}
\hfill
\includegraphics[width=0.49\textwidth]{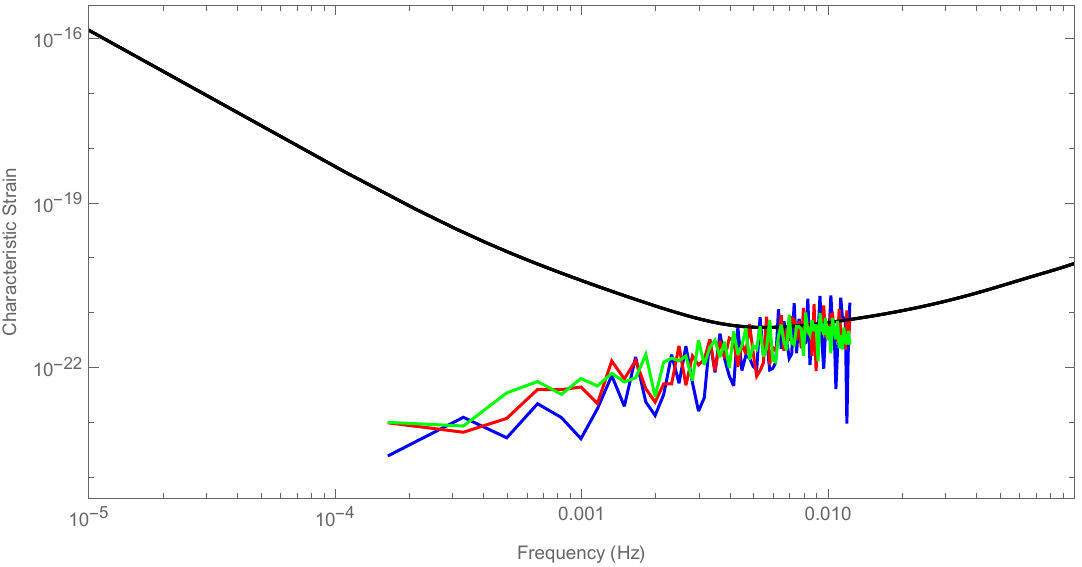}
\caption{Characteristic strain $h_c(f)$ of the gravitational-wave signals for the $q=3/2$, $(z,w,v)=(2,1,1)$ periodic orbit (left panel) and the $q=4/3$, $(z,w,v)=(3,1,1)$ periodic orbit (right panel). The curves correspond to $\alpha/M=0.10$ (blue), $0.11$ (red), and $0.12$ (green), with $a/M=0.3$ and $L/M=2.8$ fixed.
The black curve represents the LISA sensitivity curve.}
\label{fig:characteristic-strain}
\end{figure*}
Figure~\ref{fig:characteristic-strain} shows the characteristic strain for the $q=3/2$ and $q=4/3$ periodic orbits together with
the LISA sensitivity curve. In both cases, the signal is concentrated mainly in the millihertz frequency band, as expected for an EMRI involving a supermassive black hole. The PFDM parameter modifies the frequency-dependent structure of the characteristic strain, with the locations and amplitudes of its prominent features varying with $\alpha/M$. These changes originate from the modification of the underlying geodesic motion, which alters the radial and azimuthal orbital frequencies and thereby redistributes the gravitational-wave signal in frequency space. For the values of $\alpha/M$ examined here, the characteristic strain does not exhibit a simple monotonic trend with increasing $\alpha/M$, since the PFDM contribution affects several aspects of the orbital motion simultaneously. The two periodic-orbit families, $q=3/2$ and $q=4/3$, also exhibit distinct characteristic-strain profiles, reflecting the differences in their underlying orbital dynamics and topology. Thus, the frequency-domain gravitational-wave signal retains information about the corresponding periodic motion. For both orbital configurations, portions of the characteristic-strain curves approach and, at several frequencies, locally reach or exceed the LISA sensitivity curve. This indicates that the gravitational-wave signals considered here lie in a frequency and strain range relevant to LISA observations of EMRIs~\cite{LISA:2017pwj,Gair:2017ynp,Babak:2017tow}. A quantitative assessment of detectability, however, would require a signal-to-noise-ratio analysis together with the adopted observation time and detector response.
\section{Conclusions}
\label{sec:conclusion}

In this work, we have examined how a surrounding perfect fluid dark
matter (PFDM) distribution influences the strong-field dynamics of a
rotating black hole and how the resulting modifications of the geodesic
motion are carried into the associated gravitational radiation. By
combining the analysis of characteristic circular orbits, rational
periodic trajectories, and their gravitational-wave signals, the present
study provides a unified description of the dynamical and radiative
consequences of the PFDM environment in a rotating spacetime.

The analysis of equatorial timelike motion shows that the PFDM
contribution modifies the locations and conserved orbital quantities of
both prograde and retrograde circular motion. As the PFDM parameter
increases, the marginally bound orbit and the innermost stable circular
orbit shift toward smaller radii in both branches. The angular-momentum
behavior remains different for the two orbital orientations: the
positive specific angular momenta of prograde motion decrease, whereas
the retrograde specific angular momenta increase from more negative to
less negative values. At the fixed spin $a/M=0.3$, the PFDM
contribution therefore changes the quantitative properties of both
branches while preserving the characteristic prograde--retrograde
asymmetry of the rotating spacetime. At the same time,
the specific energy of the innermost stable circular orbit increases
with the PFDM parameter. These results show that the influence of the
surrounding matter distribution is not limited to a simple radial
displacement of the characteristic orbits; it also changes the
conserved quantities required to support circular motion while
preserving the characteristic prograde--retrograde asymmetry generated
by frame dragging.

A complementary picture emerges from the periodic-orbit sector. The
equatorial bound motion was organized through the rational frequency
ratio $q=\frac{\omega_{\phi}}{\omega_{r}}-1
 = w+\frac{v}{z}$, together with the $(z,w,v)$ classification. For fixed spin and specific
angular momentum, the energy associated with a given periodic-orbit
family increases with the PFDM parameter, while at fixed PFDM parameter
it also increases with $q$. More importantly, the dark-matter
contribution changes the spatial configuration of a periodic trajectory
without altering the prescribed number of leaves and whirls. The
topological classification of the orbit is therefore retained, whereas
its energetic and geometrical realization depends on the surrounding
PFDM environment. This distinction is useful because it separates the
topological information carried by the rational frequency relation from
the quantitative changes induced by the background geometry.

The gravitational-wave analysis provides the radiative counterpart of
these orbital modifications. Within the numerical-kludge treatment, the
PFDM parameter affects the waveform indirectly through its influence on
the geodesic trajectory and hence on the position, velocity, and
acceleration of the orbiting compact object. The resulting
$h_{+}$ and $h_{\times}$ polarizations retain the characteristic
temporal structure associated with zoom--whirl motion. Motion at larger
orbital radii produces comparatively smoother portions of the signal,
whereas passages through the strong-field region are accompanied by
more rapid oscillatory behavior. Varying the PFDM parameter changes the
timing and separation of these features, while their amplitudes do not
follow a simple monotonic trend. The waveform therefore reflects not
only the orbital topology but also the PFDM-induced modification of the
underlying dynamical timescales. The frequency-domain behavior is further characterized through the
corresponding characteristic strain. For the representative
extreme-mass-ratio system considered here, the characteristic strain
is concentrated mainly in the millihertz frequency band and extends
into the frequency-strain region relevant to LISA observations.
The PFDM parameter modifies the frequency-dependent structure of the
characteristic strain through its influence on the underlying orbital
dynamics. Because the present treatment does not include a full
detector response or signal-to-noise-ratio calculation, the comparison
with the LISA sensitivity curve should be regarded as an indication of
observational relevance rather than as a quantitative detectability
forecast.

The overall picture that emerges is that the PFDM contribution acts
consistently across the orbital and radiative sectors. A modification
of the background geometry first changes the characteristic circular
orbits and the conserved quantities governing timelike motion; it then
alters the energy and spatial configuration of periodic trajectories;
finally, these dynamical changes are transferred to the temporal
waveforms and the corresponding characteristic strain. Periodic motion
therefore provides a natural bridge between environmental modifications
of the strong-field geometry and their possible radiative
manifestations in extreme-mass-ratio systems.

The present analysis is based on equatorial geodesic motion and a
leading-order numerical-kludge description of the emitted radiation.
A more complete treatment should incorporate radiation reaction,
self-force effects, higher multipolar contributions, and generic
nonequatorial motion. Combining these ingredients with a realistic
detector response and a signal-to-noise-ratio analysis would make it
possible to determine more quantitatively whether the PFDM-induced
modifications identified here can be distinguished from changes in the
intrinsic parameters of the central black hole. Such extensions would
provide a natural next step toward assessing the extent to which
gravitational waves from extreme-mass-ratio systems can probe the
environment surrounding rotating black holes.

\section*{Acknowledgement}

 This work is supported by the National Natural Science Foundation of China under Grants No.~12675080, No.~12542053, and No.~12275238, the Zhejiang Provincial Natural Science Foundation of China under Grants No.~LR21A050001 and No.~LY20A050002, the National Key Research and Development Program of China under Grant No. 2020YFC2201503, and the Fundamental Research Funds for the Provincial Universities of Zhejiang in China under Grant No.~RF-A2019015. 

\bibliography{ref_bh}

\end{document}